\documentclass[pra,a4paper,preprintnumbers,floatfix,superscriptaddress,twocolumn,showpacs,notitlepage,longbibliography, aps]{revtex4-2}

\usepackage[utf8]{inputenc}
\usepackage[T1]{fontenc}
\usepackage{amsmath, amsthm, amssymb,amsfonts,mathbbol,amstext}
\usepackage{graphicx}
\usepackage{dcolumn}
\usepackage{bm}
\usepackage{bbm}
\usepackage{hyperref}
\usepackage{mathtools}
\usepackage{comment}
\usepackage{color}
\usepackage{multirow}
\usepackage{pgf,tikz}
\usepackage{mathrsfs}
\usepackage{physics}
\usetikzlibrary{arrows}
\usepackage{soul,xcolor}
\usepackage{adjustbox}
\usepackage{dsfont}

\begin{document}

\title{Rugosity as a measurement-induced relative entropy}

\author{Lucas Chibebe C\'eleri\href{https://orcid.org/0000-0001-5120-8176}{\includegraphics[scale=0.05]{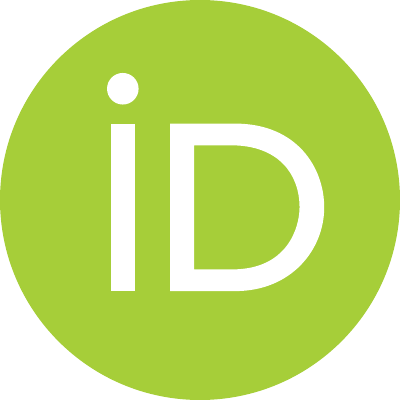}}}
\email{lucas@qpequi.com}
\affiliation{QPequi Group, Institute of Physics, Federal University of Goi\'as, Goi\^ania, Goi\'as, 74.690-900, Brazil}
\affiliation{Instituto de Física de São Carlos, Universidade de São Paulo, CP 369, 13560-970, São Carlos, SP, Brasil}

\author{Diogo O. Soares-Pinto}
\affiliation{Instituto de Física de São Carlos, Universidade de São Paulo, CP 369, 13560-970, São Carlos, SP, Brasil}

\begin{abstract}
Quantum-state rugosity quantifies the departure of a quantum state from the flat superposition associated with a prescribed basis. We show that rugosity equals the classical relative entropy generated by the binary flat-state measurement. This identity gives two hypothesis-testing interpretations. Rugosity is the asymptotic discrimination exponent under repeated flat-state measurements and the exact finite-copy exponent under unrestricted quantum testing with zero type-I error. For smoothly parametrized states, we relate its local variation to the Fisher information accessible through the prescribed measurement. Using the local expansion of quantum relative entropy, we derive a bound on rugosity sensitivity in terms of the entropy-production susceptibility of a subsequent thermal reset. This susceptibility is given by the Bogoliubov-Kubo-Mori Fisher information. Finally, we formulate a measurement-dependent quantum speed limit whose endpoint distinguishability is determined by rugosity. These results connect quantum-state rugosity with statistical discrimination, thermal reset, and the geometry of experimentally accessible probability distributions.
\end{abstract}

\maketitle  

\section{Introduction}

Quantum mechanics allows for the existence of states exhibiting intricate structures in a given representation, ranging from interference fringes in phase space to highly oscillatory wave functions in Hilbert space. Such structures often encode genuinely quantum features and have been linked to a variety of physical phenomena, including quantum coherence~\cite{Baumgratz2014}, nonclassicality~\cite{Vogel2000}, chaos~\cite{Haake2019}, metrology~\cite{Giovannetti2011}, and quantum technologies~\cite{Nielsen2011}. Motivated by these observations, the notion of quantum-state texture has recently emerged as a framework to quantify the degree of structural complexity exhibited by a quantum state with respect to a prescribed basis~\cite{Parisio2024}. Among the first and most fundamental measures introduced in this context is rugosity, which quantifies the departure of a quantum state from the completely flat superposition defined by the chosen basis~\cite{Parisio2024}. Since its introduction, the subject has developed rapidly, leading to several generalized texture measures based on different distance functionals and resource-theoretic constructions~\cite{Wang2025,Chen2026,Cao2026}. These developments have considerably expanded the mathematical framework of quantum-state texture and established it as a promising direction in quantum information theory.

Although rugosity is directly measurable and has been applied to quantum-gate identification~\cite{Parisio2024}, its interpretation in terms of statistical hypothesis testing deserves further investigation. Existing studies have developed alternative texture measures~\cite{Wang2025}, resource-theoretic constructions~\cite{Parisio2024,Greenwood2026}, and geometric and information-theoretic connections~\cite{Muthuganesan2026}. Here, we focus specifically on the original logarithmic rugosity and identify the discrimination tasks that it quantifies. We then use this interpretation to derive constraints on its local sensitivity and dynamical evolution.

Statistical distinguishability constitutes one of the cornerstones of modern information theory~\cite{Cover2006}. Relative entropy quantifies the distinguishability between probability distributions and governs the asymptotic performance of hypothesis testing through Stein's lemma~\cite{Cover2006,Hiai1991}. Its local curvature gives rise to the Fisher information~\cite{Nagaoka2007}, which underlies statistical estimation~\cite{Helstrom1969,Holevo2011}, information geometry~\cite{Nagaoka2007}, thermodynamic uncertainty relations~\cite{Micadei2013,Barato2015,Horowitz2020}, and quantum speed limits~\cite{Taddei2013,Pires2016,Pires2021}. In the quantum setting, these concepts are unified by the quantum relative entropy and its associated monotone Riemannian metrics~\cite{Petz1996}. Together, they provide a geometric description of statistical inference and nonequilibrium processes that has found applications across quantum information~\cite{Nielsen2011}, statistical physics~\cite{Kawai2007} and quantum thermodynamics~\cite{Esposito2009}.

In this work, we identify rugosity with the relative entropy between the outcome distributions obtained by applying the binary flat-state measurement to the flat state and an arbitrary quantum state. This establishes both a measurement-restricted asymptotic interpretation and an exact finite-copy interpretation under the requirement of zero type-I error. For parametrized states, we express the Fisher information of the measured distributions in terms of rugosity and its derivative.

We subsequently consider the entropy produced when an encoded state relaxes to a fixed thermal reference state. The local expansion of quantum relative entropy and its monotonicity under measurement yield a bound on rugosity sensitivity in terms of the entropy-production susceptibility of the reset. Finally, we derive a quantum speed limit using the statistical distance between the measured distributions. Throughout, the relevant statistical geometry is that of the prescribed measurement outcomes, while the BKM information connects the local sensitivity bound to thermal reset.

We use natural logarithms and set \(k_B=\hbar=1\). Entropies and entropy production are therefore dimensionless.

\section{Hypothesis testing}

We start by showing that rugosity admits a natural information-theoretic interpretation in terms of statistical distinguishability and hypothesis testing.

Let $\mathcal H_d$ be a finite-dimensional Hilbert space of dimension $d$, and let $\{\ket{\epsilon_m}\}_{m=1}^{d}$ be a fixed orthonormal basis. Associated with this basis, we define the flat state
\begin{equation}
\ket{\omega} = \frac{1}{\sqrt d} \sum_{m=1}^{d}\ket{\epsilon_m}.
\label{eq:flat_state}
\end{equation}
The definition of a flat state assumes a fixed phase
convention for the prescribed basis, which also determines
the corresponding measurement projector. Given a density operator $\rho$, the rugosity of $\rho$ with respect to this basis is defined as~\cite{Parisio2024}
\begin{equation}
R(\rho) = -\ln\Tr(\rho\Pi_\omega),
\label{eq:rugosity}
\end{equation}
with $\Pi_\omega=\ket{\omega}\bra{\omega}$ being the projector onto the flat state. The quantity $p_\omega(\rho) = \Tr(\rho\Pi_\omega)$ is the probability that the state $\rho$ passes the binary test associated with the flat state. We allow $R(\rho)=+\infty$ when $\operatorname{Tr}(\rho\Pi_\omega)=0$. Differential expressions involving rugosity will be used on regions where $0<\operatorname{Tr}(\rho\Pi_\omega)<1$, with boundary values considered through limits when these exist. The prescribed basis and its phase convention, and hence $\Pi_\omega$, remain fixed throughout each parametrized or dynamical process.

Consider the binary projective measurement $\mathcal M_\omega = \{\Pi_\omega,\Pi_\omega^c\}$, with $\Pi_\omega^c = \mathbb I-\Pi_\omega$. When this measurement is performed on $\rho$, it produces the probability distribution $P_\rho^{(\omega)} = \bigl(p_\omega,1-p_\omega\bigr)$. If the same measurement is performed on the flat state $\Pi_\omega$, the deterministic distribution is obtained $P_\omega=(1,0)$.

The classical relative entropy between two probability distributions $p=\{p_x\}$ and $q=\{q_x\}$ is
\[
K(p\|q) = \sum_x p_x \ln\frac{p_x}{q_x}.
\]
From this definition, it follows immediately that the rugosity can be written as
\begin{equation}
R(\rho) = K\left(P_\omega\|P_\rho^{(\omega)}\right).
\label{eq:rugosity_classical}
\end{equation}

More generally, given two quantum states $\sigma$ and $\rho$ and a fixed POVM $\mathcal M=\{M_x\}$, the measurement induces the probability distributions $p_x=\Tr(\sigma M_x)$ and $q_x=\Tr(\rho M_x)$, from which one defines the measurement-induced relative entropy
\begin{equation}
D_{\mathcal M}(\sigma\|\rho) = K(p\|q).
\label{eq:DM}
\end{equation}
The quantity $D(\sigma\|\rho)=\Tr[\sigma(\ln\sigma-\ln\rho)]$ is the quantum relative entropy. The subscript $\mathcal M$ indicates that the distinguishability is evaluated after both states are subjected to measurement $\mathcal M$.

Equation~\eqref{eq:rugosity_classical} is therefore equivalent to
\begin{equation}
R(\rho) = D_{\mathcal M_\omega}\bigl(\Pi_\omega\|\rho\bigr).
\label{eq:rugosity_DM}
\end{equation}
Interestingly, rugosity also coincides with the order-zero Petz Rényi relative entropy, $R(\rho)=D_0(\Pi_\omega\Vert\rho)$, where $D_0(\sigma\Vert\rho)=-\ln\Tr(\Pi_\sigma\rho)$ and $\Pi_\sigma$ projects onto the support of $\sigma$. Thus, rugosity can be interpreted as the distinguishability between $\rho$ and the flat state when the only measurement allowed is the binary flat state test.

One may maximize Eq.~\eqref{eq:DM} on all POVMs. This leads to the measured relative entropy~\cite{Hiai1991,Donald1986,Petz1986}
\begin{equation}
D_{\mathrm{mea}} (\sigma\|\rho) = \sup_{\mathcal M}D_{\mathcal M}(\sigma\|\rho).
\end{equation}
The measured relative entropy quantifies the largest classical distinguishability that can be extracted from the pair of quantum states by means of a measurement.

Therefore, we have the following hierarchy
\begin{equation}
R(\rho) \leq D_{\mathrm{mea}}\bigl(\Pi_\omega\|\rho\bigr) \leq D\bigl(\Pi_\omega\|\rho\bigr) = -\bra{\omega}\ln\rho\ket{\omega}.
\label{eq:rugosity_hierarchy}
\end{equation}
The final expression is finite when $\ket{\omega}$ belongs to the support of $\rho$; otherwise, $D(\Pi_\omega\Vert\rho)=+\infty$. The first inequality follows directly from the maximization over the measurements. The second follows from the monotonicity of the quantum relative entropy under quantum channels, taking the measurement map as a particular quantum channel.

The hypothesis-testing interpretation now follows naturally. We distinguish the hypothesis that the prepared state is $\Pi_\omega$ from the alternative that it is $\rho$. Suppose that only measurement $\mathcal M_\omega$ is available. Repeating this measurement in many independently prepared copies produces the probability distributions $P_\omega$ and $P_\rho^{(\omega)}$. The discrimination problem then reduces to a classical hypothesis-testing problem between these two distributions.

The type-I error is the probability of rejecting the flat-state hypothesis when it is true, while the type-II error is the probability of accepting it when the prepared state is $\rho$. Under repeated applications of the flat-state measurement, classical Stein's lemma identifies $K(P_\omega\|P_\rho^{(\omega)})$ as the optimal asymptotic decay exponent of the type-II error when the type-I error is bounded by a fixed tolerance $0<\epsilon<1$. By Eq.~\eqref{eq:rugosity_classical}, this exponent is precisely rugosity.

In fact, a stronger finite-copy statement can be established. For $n$ independently prepared copies, consider the binary test $\{\Pi_\omega^{\otimes n},\mathbb{I}-\Pi_\omega^{\otimes n}\}$. Taking the flat state as the null hypothesis, this test has a vanishing type-I error, while the type-II error is $\beta_n=p_\omega^n=e^{-nR(\rho)}$. Consequently,
\begin{equation}
    R(\rho)=-\frac{1}{n}\ln\beta_n.
    \label{eq:typeIIerror}
\end{equation}
The test above never rejects the flat state when it is present and minimizes the type-II error among all quantum tests with this property. Indeed, accepting the flat state with certainty forces the acceptance operator to act as the identity on its one-dimensional support. Positivity implies that its action on the orthogonal complement can only add a nonnegative contribution to the type-II error. Equation~\eqref{eq:typeIIerror} therefore gives a second operational interpretation of rugosity as the exact finite-copy discrimination exponent when the type-I error must vanish, without restricting the available quantum measurements.

Rugosity does not generally equal the measured or quantum relative entropy in Eq.~\eqref{eq:rugosity_hierarchy}. It quantifies distinguishability under the prescribed flat-state measurement and also determines the optimal finite-copy error exponent when the type-I error is required to vanish exactly.

The relation between rugosity and relative entropy will be useful in the following sections, where it allows us to connect rugosity with Fisher information, thermodynamic sensitivity bounds, and quantum speed limits.

\section{Thermodynamic sensitivity bound}

The information-theoretic interpretation of rugosity developed in the previous section naturally leads to a connection between rugosity and the entropy produced during thermal relaxation after parameter encoding. The basic observation is that relative entropy plays a dual role in information theory and thermodynamics. On the one hand, relative entropy quantifies statistical distinguishability. On the other hand, it appears in fundamental expressions for dissipation and entropy production. Since Fisher information emerges as the local curvature of relative entropy, the relation between rugosity and relative entropy naturally leads to a connection between rugosity and thermodynamic sensitivity bounds.

Consider a family of states $\rho_\phi$ that smoothly depend on a parameter $\phi$. The parameter may represent time, an external field, a coupling strength, a driving amplitude, a phase, or any thermodynamic control parameter. Associated with this family, we define $p_\phi=\Tr(\rho_\phi\Pi_\omega)$, so that the rugosity becomes $R(\phi)=-\ln p_\phi$. The flat-state measurement $\mathcal M_\omega$ produces the binary probability distribution $P_\phi^{(\omega)} = (p_\phi,1-p_\phi)$. For $0<p_\phi<1$, the classical Fisher information of this Bernoulli distribution takes the form
\begin{equation}
F_\phi^{(\omega)} = \frac{(\partial_\phi R_\phi)^2}{e^{R_\phi}-1},
\label{eq:rugosity_fisher}
\end{equation}
where we used $p_\phi=e^{-R_\phi}$. 

Equation~\eqref{eq:rugosity_fisher} relates the magnitude of the rugosity derivative to its local value and the Fisher information associated with the flat-state measurement.

Let $\rho_0=e^{-\beta H_0}/Z_0$ denote the initial equilibrium state, and let $\rho_\phi$ be the nonequilibrium state obtained after the encoding protocol. We assume $\rho_{\phi=0}=\rho_0$ and finite temperature, with $\beta$ and $H_0$ fixed throughout the family of reset processes. To quantify the thermodynamic departure from equilibrium, we consider a subsequent thermal reset, during which the system relaxes to $\rho_0$ in contact with a thermal reservoir at inverse temperature $\beta$. Assuming an initially uncorrelated thermal reservoir, weak coupling with negligible interaction-energy contributions, complete thermalization, and a fixed Hamiltonian $H_0$ with no work performed during reset, the total entropy production associated with this process is~\cite{Esposito2011}
\begin{equation}
\Sigma_\phi = D(\rho_\phi\|\rho_0),
\label{eq:entropy_production}
\end{equation}
which represents the entropy produced during the subsequent thermal reset.

As shown in Appendix~\ref{app:expansion}, the second-order expansion of the quantum relative entropy around equilibrium is
\begin{equation}
\Sigma_\phi = \frac{\phi^2}{2} F_{\mathrm{BKM}} + O(\phi^3),
\label{eq:SigmaExpansion}
\end{equation}
where $F_{\mathrm{BKM}}$ is the Bogoliubov-Kubo-Mori Fisher information associated with the family $\rho_\phi$, evaluated at $\phi=0$~\cite{Petz1996}. Since the BKM metric is monotone under completely positive trace-preserving maps and reduces to the classical Fisher information after a measurement, the Fisher information associated with the flat-state measurement satisfies
\begin{equation}
    F_0^{(\omega)}\leq F_{\mathrm{BKM}}.
    \label{eq:FisherHierarchy}
\end{equation}

Introducing the entropy-production susceptibility
\begin{equation}
\chi_\Sigma = \left. \frac{\partial^2\Sigma_\phi} {\partial\phi^2} \right|_{\phi=0},
\label{eq:SigmaSusceptibility}
\end{equation}
Eq.~\eqref{eq:SigmaExpansion} immediately implies $\chi_\Sigma = F_{\mathrm{BKM}}$. Combining this with Eqs.~\eqref{eq:rugosity_fisher} and \eqref{eq:FisherHierarchy} we obtain
\begin{equation}
\left.\left(\partial_\phi R\right)^2\right|_{\phi=0} \leq
\left(e^{R_{\phi=0}}-1\right)\chi_\Sigma.
\label{eq:RugosityBound2}
\end{equation}
Equation~\eqref{eq:RugosityBound2} establishes a direct connection between rugosity and nonequilibrium thermodynamics. It shows that the local sensitivity of rugosity is bounded by the susceptibility of the entropy produced upon restoring equilibrium. Thus, the bound quantifies how the sensitivity of rugosity is constrained by the thermodynamic distance from the reference equilibrium state, which is characterized by the entropy produced upon restoring equilibrium.

The relation established in Eq.~\eqref{eq:rugosity_fisher} extends beyond the thermodynamic setting. Since it connects the sensitivity of rugosity to the Fisher information associated with the flat-state measurement, it also provides a direct link between rugosity and the geometry of quantum-state evolution. In the next section, we exploit this connection to derive a quantum speed limit expressed in terms of rugosity.

\section{Operational quantum speed limit}

We now derive a speed limit directly from the statistical geometry induced by the flat-state measurement.

Let the state evolve in time as $\rho_t$, and define $p_t=\Tr(\rho_t\Pi_\omega)$ and $R_t=-\ln p_t$. The flat-state measurement produces the binary probability distribution $P_t^{(\omega)}=(p_t,1-p_t)$. The corresponding classical Fisher information is
\begin{equation}
F_t^{(\omega)} = \frac{\dot p_t^{\,2}}{p_t(1-p_t)} = \frac{\dot R_t^{\,2}}{e^{R_t}-1}.
\label{eq:rugosity_time_fisher}
\end{equation}
These expressions involving rugosity are understood for
$0<p_t<1$, with boundary cases treated through
appropriate limits whenever they exist. The statistical distance between two probability distributions is given by the classical Bhattacharyya angle~\cite{Braunstein1994}, which takes the form
\[
\mathcal L_\omega(p_0,p_\tau) = \arccos\left[\sqrt{p_0p_\tau} + \sqrt{(1-p_0)(1-p_\tau)}\right]
\]
for a Bernoulli distribution. Using $p_t=e^{-R_t}$, we can write this distance in terms of rugosity
\[
\mathcal L_\omega(p_0,p_\tau) = \arccos\left[e^{-\frac{R_0+R_\tau}{2}} + \sqrt{\left(1-e^{-R_0}\right) \left(1-e^{-R_\tau}\right)}\right].
\]

In order to derive the quantum speed limit, we observe that the infinitesimal form of the Bhattacharyya angle is
\[
\dd s_\omega^2 = \frac{1}{4}F_t^{(\omega)}\dd t^2.
\]
Therefore, the length of the path followed by the measured probability distribution is
\[
\ell_\omega[0,\tau] = \frac{1}{2} \int_0^\tau \sqrt{F_t^{(\omega)}}\,\dd t = \frac{1}{2}\int_0^\tau\frac{|\dot R_t|}{\sqrt{e^{R_t}-1}}\,\dd t .
\]
Since the geodesic distance cannot exceed the length of the path,
\begin{equation}
\mathcal L_\omega(p_0,p_\tau) \leq \ell_\omega[0,\tau].
\label{eq:length_distance}
\end{equation}
This geometric inequality relates the endpoint distinguishability to the length of the trajectory of the measured probability distribution.

For unitary dynamics generated by a Hamiltonian $H_t$, the probability satisfies $\dot p_t = i\Tr\left(\rho_t[H_t,\Pi_\omega]\right)$. The instantaneous energy uncertainty is
\begin{equation}
    \Delta H_t  = \sqrt{\operatorname{Tr}(\rho_tH_t^2) -\left[\operatorname{Tr} (\rho_tH_t)\right]^2}.
    \label{eq_energy_uncertainty}
\end{equation}
Using the Robertson uncertainty relation with the observable $\Pi_\omega$, one obtains $\abs{\dot p_t} \leq 2\Delta H_t\,\Delta\Pi_\omega$. Since $\Pi_\omega$ is a projector, $(\Delta\Pi_\omega)^2 = p_t(1-p_t)$. Hence,
\begin{equation}
F_t^{(\omega)} = \frac{\dot p_t^{\,2}}{p_t(1-p_t)} \leq 4(\Delta H_t)^2.
\label{eq:restricted_fisher_energy}
\end{equation}
In terms of rugosity, this becomes
\begin{equation}
\frac{\abs{\dot R_t}} {\sqrt{e^{R_t}-1}} \leq 2\Delta H_t .
\label{eq:rugosity_rate_energy}
\end{equation}

Combining Eqs.~\eqref{eq:length_distance} and \eqref{eq:restricted_fisher_energy}, we obtain
\begin{equation}
\mathcal L_\omega(p_0,p_\tau) \leq \int_0^\tau \Delta H_t\,dt .
\label{eq:measurement_qsl_general}
\end{equation}
Therefore,
\begin{equation}
\tau \geq \frac{\mathcal L_\omega(p_0,p_\tau)} {\overline{\Delta H}},
\label{eq:measurement_qsl}
\end{equation}
where
\begin{equation}
\overline{\Delta H} = \frac{1}{\tau} \int_0^\tau \Delta H_t\,\dd t .
\end{equation}
For a time-independent Hamiltonian, $\Delta H_t=\Delta H$, and the bound reduces to
\begin{equation}
\tau \geq \frac{\mathcal L_\omega(p_0,p_\tau)}{\Delta H}.
\label{eq:measurement_qsl_time_independent}
\end{equation}
It is worth noting that the present bound is closely related to the conventional Mandelstam-Tamm quantum speed limit. In particular, when the initial state is the flat state, $\rho_0=\Pi_\omega$, we have $p_0=1$ and $\mathcal{L}_\omega(p_0,p_\tau)=\arccos\sqrt{p_\tau}$. For unitary dynamics, this coincides with the Bures angle between the initial and final states, recovering the Mandelstam-Tamm bound. For general initial states, however, the present bound involves only the distinguishability accessible through the prescribed flat-state measurement, rather than the full quantum-state distinguishability.

This result provides an operational quantum speed limit whose endpoint distinguishability is expressed entirely in terms of rugosity.. The conventional quantum speed limit is formulated in terms of the distinguishability between quantum states and therefore involves the Bures geometry. The present construction follows a different philosophy. Since rugosity is defined through a fixed measurement, the relevant geometry is not the geometry of the full quantum state space, but that of the statistical manifold generated by the measurement outcomes. The resulting speed limit therefore provides a lower bound on the evolution time required to produce a given change in the probability distribution accessible through the flat-state measurement.

\section{Conclusions}

In this work, we showed that rugosity admits a natural information-theoretic interpretation as a measurement-induced relative entropy. This identification provides rugosity with a direct operational meaning: it quantifies the distinguishability between a quantum state and the corresponding flat state when the available information is restricted to the binary flat state measurement. Rather than being simply a measure of quantum-state texture, rugosity becomes a particular instance of measurement-restricted statistical distinguishability.

This perspective naturally connects rugosity with several seemingly unrelated concepts. The magnitude of the rugosity derivative is determined by its local value and the Fisher information associated with the flat-state measurement. The statistical geometry of the resulting probability distributions therefore provides a common framework for the sensitivity and speed-limit bounds derived here. In particular, we showed that the local variation of rugosity is constrained by the entropy-production susceptibility of the subsequent thermal reset, while its dynamical evolution satisfies an operational quantum speed limit determined by the statistical geometry of the measurement outcomes. These results connect hypothesis testing, the entropy produced during thermal reset, and quantum dynamics through the distinguishability accessible to the prescribed flat-state measurement.

The present work also suggests several directions for future investigation. A natural direction is to investigate analogous discrimination tasks and thermodynamic sensitivity bounds for other experimentally prescribed POVMs. It would also be interesting to investigate the geometry of the resulting restricted statistical manifolds and their relation to the intrinsic geometry of quantum state space. Finally, the connection established here between rugosity, Fisher information, and entropy production suggests the possibility of developing measurement-restricted thermodynamic sensitivity bounds and finite-time bounds on texture generation, thereby extending recent information-theoretic formulations of nonequilibrium thermodynamics~\cite{Nicholson2020} to experimentally constrained settings.

\begin{acknowledgments}
LCC acknowledges support from the National Council for Scientific and Technological Development (CNPq) through grant 308065/2022-0, the National Institute of Science and Technology for Applied Quantum Computing through CNPq grant 408884/2024-0, Goiás State Research Foundation (FAPEG) through grant 202510267001843, and São Paulo State Research Foundation (FAPESP) through grant 2025/23726-4. DOSP acknowledges support from the National Council for Scientific and Technological Development (CNPq) under Grant No. 305766/2026-0. 
\end{acknowledgments}

\appendix

\section{Expansion of the relative entropy}
\label{app:expansion}

Let $\{\rho_\phi\}$ be a smooth family of full-rank density operators depending on a real parameter $\phi$. Expanding around $\phi=0$, we obtain
\begin{equation}
\rho_\phi = \left.\rho_\phi\right|_{\phi=0} + \phi\left.\dv{\rho_\phi}{\phi}\right|_{\phi=0} + \frac{\phi^2}{2}\left.\dv[2]{\rho_\phi}{\phi}\right|_{\phi=0} + O(\phi^3),
\end{equation}
Our goal is to expand the quantum relative entropy $f_\phi = D(\rho_\phi\|\rho_0)$. The first term of the expansion vanishes, $f_{\phi=0} = D(\rho_0\|\rho_0) = 0$.

From here on, to simplify notation we write $\rho_a \equiv \left.\rho_\phi\right|_{\phi=a}$ and
\[
\dot\rho_{a} \equiv \left.\dv{\rho_\phi}{\phi}\right|_{\phi=a},
\]
with double dots for the second derivative. To determine the first nontrivial contribution, we differentiate with respect to $\phi$. Using the identity
\begin{equation}
\dv{}{\phi} \Tr(\rho_\phi\ln\rho_\phi) = \Tr(\dot\rho_\phi\ln\rho_\phi) + \Tr(\dot\rho_\phi),
\end{equation}
together with $\Tr(\dot\rho_\phi)=0$ from the normalization of the density operator, one finds
\begin{equation}
\left.\dv{f_\phi}{\phi}\right|_{\phi=0} = \Tr\left[\dot\rho_0 \left(\ln\rho_0 - \ln\rho_0\right)\right]= 0.
\end{equation}
Thus, the relative entropy has no linear contribution around $\phi=0$.

Differentiating once more we obtain
\begin{equation}
\left.\dv[2]{f_\phi}{\phi}\right|_{\phi=0} = \Tr\left[\dot\rho_0\left. \dv{}{\phi}\ln\rho_\phi\right|_{\phi=0}\right].
\label{eq:fpp}
\end{equation}
In order to compute the derivative inside the trace, we use the integral representation of the logarithm of a full rank operator $A$
\[
\ln A = \int_{0}^{\infty}\dd s\left(\frac{\mathbb{1}}{1+s} - \left[A + s\mathbb{1}\right]^{-1}\right)
\]
and the derivative of the inverse operator
\[
\dv{A^{-1}}{\phi} = -A^{-1}\dv{A}{\phi}A^{-1}.
\]
The result is the Fréchet derivative~\cite{Bhatia2009}
\[
\left.\dv{}{\phi}\ln\rho_\phi\right|_{\phi=0} = \int_0^\infty \dd s\,(\rho_0+s\mathbb I)^{-1}\dot\rho_0(\rho_0+s\mathbb I)^{-1}.
\]
Substituting this equation into Eq.~\eqref{eq:fpp} leads to
\[
\left.\dv[2]{f_\phi}{\phi}\right|_{\phi=0} = \int_0^\infty \dd s\, \Tr\left[\dot\rho_0(\rho_0+s\mathbb I)^{-1}\dot\rho_0(\rho_0+s\mathbb I)^{-1}\right].
\]
 
Joining everything, the expansion of the relative entropy up to second order is given by
\begin{equation}
D(\rho_\phi\|\rho_0) = \frac{\phi^2}{2} F_{\rm BKM} + O(\phi^3),
\end{equation}
where
\begin{equation}
F_{\rm BKM} = \int_0^\infty \dd s\, \Tr\left[\dot\rho_0(\rho_0+s\mathbb I)^{-1}\dot\rho_0(\rho_0+s\mathbb I)^{-1}\right]
\label{eq:BKM}
\end{equation}
is the Bogoliubov-Kubo-Mori Fisher information~\cite{Petz1996}.

To obtain a more explicit expression, let $\rho_0 = \sum_n p_n |n\rangle\langle n|$ be the spectral decomposition of $\rho_0$. Then,
\[
(\rho_0+s\mathbb I)^{-1} = \sum_n \frac{1}{p_n+s}|n\rangle\langle n|.
\]
Defining $O_{mn} = \mel{m}{O}{n}$, Eq.~\eqref{eq:BKM} becomes
\begin{eqnarray*}
F_{\rm BKM} &=& \sum_{m,n} |(\dot\rho_0)_{mn}|^2\int_0^\infty\frac{\dd s}{(p_m+s)(p_n+s)} \\
&=& \sum_{m,n} \frac{\ln p_m-\ln p_n}
{p_m-p_n}|(\dot\rho_0)_{mn}|^2.
\end{eqnarray*}
For $p_m=p_n$, the divided difference is understood through its continuous limit,
\begin{equation}
    \lim_{y\to x} \frac{\ln x-\ln y}{x-y} =\frac{1}{x}.
\end{equation}

This quantity differs, in general, from the symmetric logarithmic derivative quantum Fisher information
\[
F_{\rm SLD} = 2 \sum_{m,n} \frac{|(\dot\rho_0)_{mn}|^2}{p_m+p_n}.
\]
The two coincide only in the commuting case, namely when $[\rho_0,\dot\rho_0]=0$. In this situation, a common eigenbasis of $\rho_0$ and $\dot\rho_0$ can be chosen, including within degenerate eigenspaces, and both expressions reduce to the classical Fisher information,
\[
F_{\rm cl} = \sum_n \frac{(\dot p_n)^2}{p_n}.
\]
Here, $\dot p_n=\langle n|\dot\rho_0|n\rangle$ in that common eigenbasis.


\end{document}